\documentclass{article}

\usepackage[english]{babel}

\usepackage[letterpaper,top=2cm,bottom=2cm,left=3cm,right=3cm,marginparwidth=1.75cm]{geometry}

\usepackage{amsmath}
\usepackage{graphicx}
\usepackage{float}
\usepackage{enumitem}
\usepackage{makecell}
\usepackage{authblk}
\usepackage{cite}
\usepackage[normalem]{ulem}
\usepackage[colorlinks=true, allcolors=blue]{hyperref}

\usepackage{pdflscape}
\usepackage{longtable}
\usepackage{booktabs}
\usepackage{siunitx}

\title{Benchmark for two-dimensional large scale coherent structures in partially magnetized E$\times$B plasmas - Community collaboration \& lessons learned}

\author[1]{Andrew T. Powis}
\author[2]{Eduardo Ahedo}
\author[3]{Alejandro \'Alvarez Laguna}
\author[6]{Nicolas Barléon}
\author[2]{Enrique Bello-Ben\'itez}
\author[4]{Lucas Beving}
\author[5]{Jean-Pierre Boeuf}
\author[6]{Guillaume Bogopolsky} 
\author[3]{Anne Bourdon}
\author[7]{Filippo Cichocki}
\author[6]{B\'en\'edicte Cuenot} 
\author[8]{Andrew Denig}
\author[9]{Zolt\'an Donk\'o}
\author[10]{Paul-Quentin Elias}
\author[2]{Miguel P. Encinar}
\author[11]{Denis Eremin}
\author[2]{Pablo Fajardo} 
\author[12]{Farbod Faraji}
\author[5]{Gwenael Fubiani}
\author[5]{Laurent Garrigues}
\author[8]{Kentaro Hara}
\author[9]{Peter Hartmann}
\author[4]{Matthew Hopkins}
\author[1]{Igor D. Kaganovich}
\author[12]{Aaron Knoll}
\author[14]{Giovanni Lapenta}
\author[15]{Thierry E. Magin} 
\author[2]{Alberto Mar\'in-Cebri\'an}
\author[2]{Mario Merino} 
\author[16]{Pierpaolo Minelli}
\author[17]{Mina Papahn Zadeh}
\author[13]{Pietro Parodi} 
\author[3]{Federico Petronio}
\author[12]{Maryam Reza}
\author[17]{Andrei I. Smolyakov}
\author[18]{Dmytro Sydorenko}
\author[16]{Francesco Taccogna}
\author[19]{Miles M. Turner}
\author[6]{Olivier Vermorel}
\author[1]{Willca Villafana}
\author[11]{Liang Xu}

\affil[1]{Princeton Plasma Physics Laboratory, Princeton, New Jersey 08540, USA}
\affil[2]{Universidad Carlos III de Madrid, 28911 Leganés, Spain}
\affil[3]{Laboratoire de Physique des Plasmas (LPP), CNRS, Sorbonne Université, École Polytechnique, Institut Polytechnique de Paris, 91120 Palaiseau, France}
\affil[4]{Sandia National Laboratories, Albuquerque, New Mexico 87132, USA}
\affil[5]{Université de Toulouse, Toulouse INP, CNRS, LAPLACE, Toulouse, France}
\affil[6]{CERFACS -- 42, avenue Gaspard Coriolis, F-31057 Toulouse, France}
\affil[7]{Nuclear (NUC) Department, ENEA C.R. Frascati, Frascati, Rome, Italy}
\affil[8]{Stanford University, 496 Lomita Mall, Stanford, California 94305, USA}
\affil[9]{HUN-REN Wigner Research Centre for Physics, Konkoly Thege Miklós str. 29-33, 1121 Budapest, Hungary}
\affil[10]{DPHY, ONERA, Université Paris-Saclay, Palaiseau, France}
\affil[11]{Ruhr University Bochum, Universitaetsstrasse 150, D-44801 Bochum, Germany}
\affil[12]{Imperial College London, Exhibition Road, London SW7 2AZ, United Kingdom}
\affil[13]{Aeronautics and Aerospace Department, von Karman Institute for Fluid Dynamics, 
Waterloosesteenweg 72, 1640 Sint-Genesius-Rode, Belgium}
\affil[14]{Department of Mathematics, KU Leuven, Celestijnenlaan 200B, 3001 Leuven, Belgium}
\affil[15]{Aero-Thermo-Mechanics Department, Université Libre de Bruxelles, Avenue F.D. Roosevelt 50, 1050 Bruxelles, Belgium}
\affil[16]{Institute for Plasma Science and Technology (ISTP), CNR, Bari, Italy}
\affil[17]{University of Saskatchewan, Saskatoon, Saskatchewan S7N 5E2, Canada}
\affil[18]{University of Alberta, Edmonton, Alberta T6G2E1, Canada}
\affil[19]{Dublin City University, Dublin, Ireland}

\begin{document}
\maketitle

\newpage

\begin{abstract}
Low-temperature plasmas are essential to both fundamental scientific research and critical industrial applications. As in many areas of science, numerical simulations have become a vital tool for uncovering new physical phenomena and guiding technological development. Code benchmarking remains crucial for verifying implementations and evaluating performance. This work continues the \textit{Landmark} benchmark initiative, a series specifically designed to support the verification of low-temperature plasma codes. In this study, seventeen simulation codes from a collaborative community of nineteen international institutions modeled a partially magnetized E$\times$B Penning discharge. The emergence of large scale coherent structures, or rotating plasma spokes, endows this configuration with an enormous range of time scales, making it particularly challenging to simulate. The codes showed excellent agreement on the rotation frequency of the spoke as well as key plasma properties, including time-averaged ion density, plasma potential, and electron temperature profiles. Achieving this level of agreement came with challenges, and we share lessons learned on how to conduct future benchmarking campaigns. Comparing code implementations, computational hardware, and simulation runtimes also revealed interesting trends, which are summarized with the aim of guiding future plasma simulation software development.
\end{abstract}

\section{Introduction}
\label{sec:intro}

The study of low-temperature plasmas (LTPs) is motivated by scientific exploration as well as their importance to numerous critical industrial technologies \cite{adamovich20222022}. LTPs exhibit complex, multiscale, and often chaotic behavior, similar to those observed in other regimes of plasma physics, with additional complexity introduced by chemical reactions and surface processes. On the applications side, LTPs are present in all of the most critical steps in microcircuit fabrication \cite{kanarik2020inside}, with additional use cases including other forms of materials processing \cite{anders2017tutorial}, power transmission technologies \cite{villafana2024establishing,goebel1993low}, and spacecraft electric propulsion \cite{charles2009plasmas,ahedo2011plasmas}. Research into LTPs therefore provides enormous benefits to both scientific and engineering communities.

As in many domains of modern science, simulations have come to form a third foundational pillar of discovery for LTPs. Plasma experiments, even in the low-temperature regime, are challenging to run and diagnose non-intrusively. Furthermore, the high-dimensional, non-linear, and multiscale nature of the governing equations often prevents analytical solution. Simulations therefore play a critical role in solving these equations under realistic conditions, and assist in interpreting experimental results. Despite their high computational cost, LTP simulations are also playing an increasingly important role in computer aided engineering of related technologies \cite{osowiecki2024achieving}. This progress has been aided by the continuous improvement of computing performance, particularly with the widespread adoption of computational accelerators \cite{juhasz2021efficient}, as well as the integration of traditional modeling techniques with machine learning \cite{faraji2025machine}. Nonetheless, computer science and algorithm design for LTPs is still a cutting edge area of research (for further discussion, see Section VII in Ref. \cite{kaganovich2020physics}).

Many of the interesting LTP phenomena and applications referenced above operate in the kinetic regime, with the plasma governed by the Boltzmann equation in combination with electrostatic or electromagnetic field equations. Due to its intuitive nature and relatively simple implementation, the particle-in-cell (PIC) algorithm has become the most common approach for modeling kinetic plasmas, including LTPs \cite{hockney2021computer,birdsall2018plasma,tajima2018computational}. The algorithm uses a mixed Eulerian/Lagrangian framework, with the fields solved on an up to three-dimensional Eulerian grid and particles evolved in a Lagrangian sense, in the up to six-dimensional phase space. Furthermore, collisions between charged species and neutrals are often implemented via Monte Carlo algorithms \cite{vahedi1995monte}. PIC is also highly parallelizable, with particles evolved in a way that is nearly embarrassingly parallel, and field solvers can take advantage of mature distributed linear algebra packages \cite{petsc-web-page,hypre,trilinos-website}. Furthermore, the particle and collision routines are amenable to GPU acceleration, making the algorithm even more appealing for use on modern heterogeneous supercomputing architectures \cite{kaganovich2020physics,juhasz2021efficient,powis2021particle,mertmann2011fine}.

Despite the relatively simple and interpretable nature of PIC, the requirement for versatile, scalable and portable software can significantly complicate code design and implementation. Therefore, as in any other field, code validation and verification play an important role in developing PIC codes for LTPs. Verification can be pursued in several ways, including through unit tests, the method of manufactured solutions \cite{tranquilli2022deterministic} and through benchmarking against other codes. In the past several years the LTP simulation community has introduced a series of benchmarks against which developers can test their software, these include the popular Turner benchmark of 2013 \cite{turner2013simulation} as well as the \textit{Landmark} benchmarks \cite{smolyakov2020anomalous,charoy20192d,villafana20212d}. The aim of this work is to extend the \textit{Landmark} series with a new benchmark configuration which models low-frequency and large-scale structures emerging in partially magnetized E$\times$B plasmas. This is pursued through two-dimensional PIC simulations of a Penning discharge cross-section, wherein a low-frequency rotating spoke is commonly observed.

Two previous benchmarks from the \textit{Landmark} series, specifically 2a \cite{charoy20192d} and 2b \cite{villafana20212d} modeled collisionless partially magnetized E$\times$B LTPs in the simplified two-dimensional geometry of a Hall thruster propulsion system. Within this regime, the magnetic field strength is sufficient to magnetically confine electrons, whereas the heavier ions are weakly-magnetized and can move freely under the influence of electric fields. Despite being confined by the applied magnetic field, it has been observed numerically, and experimentally, that electrons are eventually transported across the field lines from the cathode to the anode, even when ignoring collisions with other species. The underlying mechanism for this ``anomalous'' transport is still an active area of investigation, with numerous mechanisms being proposed. These include near-wall conductivity \cite{morozov1968effect,morozov2001theory} and a wide range of plasma waves and instabilities \cite{choueiri2001plasma,chesta2001characterization,adam04, tac09,heron13,lafleur2016theoryII,lafleur2017role,boniface2006anomalous,Taccogna_2019,charoy2020comparison,brown2023anomalous,boeuf2017tutorial}, with the true mechanism likely being a combination of effects and also dependent on the device under consideration. One important plasma mode which does not emerge within the Landmark 2 configurations is the so-called rotating plasma ``spoke'' \cite{janes1966anomalous}, a low-frequency, large-scale coherent structure observed in numerous E$\times$B experiments \cite{esipchuk1974plasma,parker2010transition,ellison2011direct,mcdonald2011rotating,ellison2012cross,griswold2012feedback,raitses2012studies,liu2014ultrahigh,sekerak2014azimuthal,cappelli2016coherent,hyatt2017hall,mazouffre2019rotating,anders2012drifting,winter2013instabilities,hecimovic2015spoke,poolcharuansin2015use,anders2017direct,hecimovic2017sputtering,panjan2017plasma,hnilica2018effect,ehiasarian2012high,brenning2013spokes,thomassen1966turbulent,sakawa1993excitation,raitses2015effects,claire2018ion,pierre2004radial,matsukuma2003spatiotemporal,jaeger2009direct,rebont2011ion,escarguel2010optical,kim2021magnetic,rodriguez2019boundary,cheon2023experimental,kim2022high,lee2023azimuthal,przybocki2025comparison,valinurov2025study,skoutnev2018fast,choi2023rotating,pushkarev2024spoke,lv2020rotating,ito2015self,held2022spoke,panjan2015non,sekerak2016mode,guglielmi2022simultaneous,yang2014propagation,hecimovic2014characteristic,hecimovic2018spokes,anders2012self,hecimovic2016spoke} and simulations \cite{powis2018scaling,carlsson2018particle,boeuf2013rotating,boeuf2014rotating,escobar2014analysing,carlsson2015multi,escobar2014global,escobar2015low,matyash2012numerical,taccogna2011three,taccogna2012physics,matyash2017investigation,kawashima2018numerical,boeuf2023physics,koshkarov2019self,tyushev2023azimuthal,xu2021direct,boeuf2023spoke,boeuf2019micro,mansour2022full,smirnov2023particle,xu2023rotating,ganta2024investigating,rokhmanenkov2019numerical,sengupta2021restructuring,sengupta2020mode,tyushev2025mode,chen2025three,poli2024comparison,boeuf2020rotating,boeuf2020new}. Numerical modeling of this mode within the Penning discharge represents an order of magnitude increase in the range of time scales when compared to the physics modeled in the \textit{Landmark 2} cases, providing a significant stress-test of code capabilities.

\begin{figure}[H]
    \centering
    \includegraphics[width=0.8\linewidth]{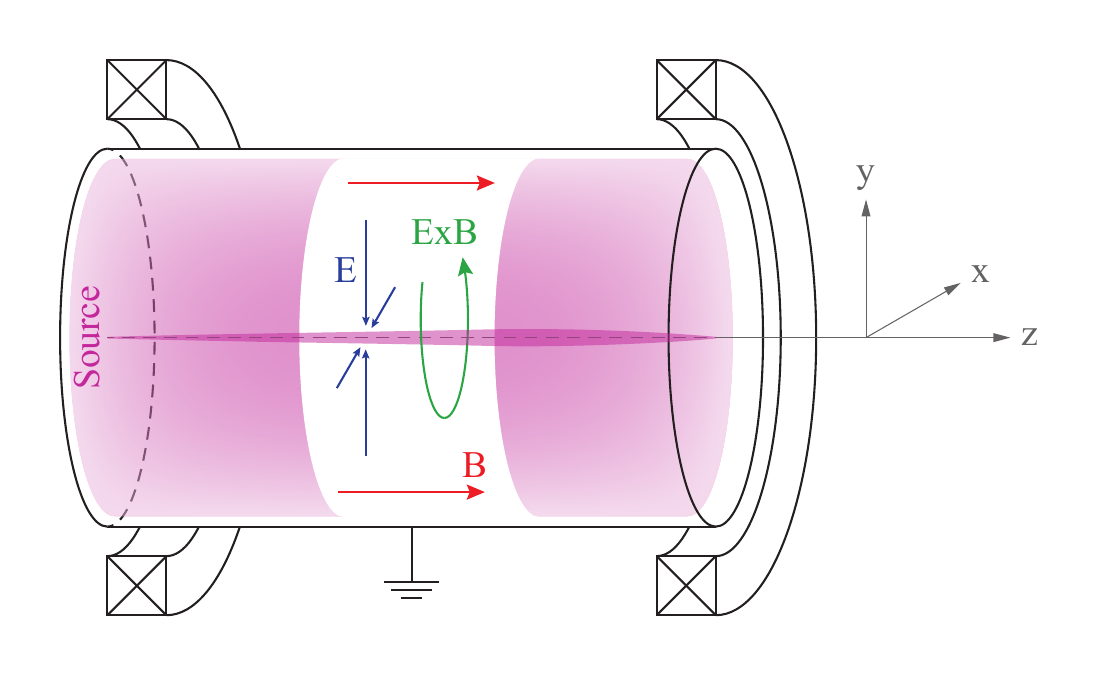}
    \caption{Illustration of a typical Penning discharge experiment with a beamline source of energetic electrons emitted from the cathode. The outer cylindrical pressure vessel is grounded, with dielectric end-plates. The approximately uniform axial magnetic field (red arrows) is sustained by electromagnets. Dynamics between the magnetized electrons and weakly magnetized ions, which are produced at the center and diffuse outwards, produces a radial ambipolar electric field (blue arrows) in the opposite direction to that of a typical ambipolar field found in un-magnetized plasmas. This results in azimuthal E$\times$B drift (green arrow), with the streaming of electrons against slowly moving ions leading to formation of the rotating spoke.}
    \label{fig:experiment}
\end{figure}

Over the past decade there has been a significant effort to develop an understanding of the spoke formation mechanism. It is generally agreed that the spoke is seeded by a gradient-drift instability (GDI), often referred to as a modified or collisionless Simon-Hoh instability (CSHI) \cite{boeuf2023spoke,xu2021direct,frias2013long,xu2023rotating,thomassen1966turbulent,sakawa1993excitation,powis2018scaling,boeuf2019micro, ito2015self}. Charge separation, resulting from the relative E$\times$B drift of magnetized electrons to the non- or weakly-magnetized ions (or possibly due to diamagnetic drift \cite{hara2022theory}) fuel linear growth and rotation in the E$\times$B direction. At long times, the spoke can coalesce into a mode or modes with amplitude comparable to the peak density within the device. Various theories have been proposed to explain the non-linear and saturated behavior of the spoke, including formation from an inverse cascade of the GDI (or possibly higher frequency) waves \cite{xu2021direct,tyushev2025mode,koshkarov2019self}. It has also been proposed that when combined with a gradient in magnetic field the spoke can transform into an ionization wave, and even reverse direction \cite{janes1966anomalous,tyushev2023azimuthal,boeuf2023spoke}. One theory also proposes that observations of the spoke are projections of a 3D helical structure \cite{chen2025three}. Just as for anomalous transport, the true answer is likely dependent on the specific system under investigation and may result from several of the above effects, as well as coupling to other waves and instabilities within the device. The structure is often strongly correlated with anomalous transport \cite{janes1966anomalous,ellison2011direct,ellison2012cross,powis2018scaling,carlsson2018particle,koshkarov2019self,tyushev2023azimuthal,boeuf2019micro}, whether due to transport being enhanced by the wave itself, or the presence of a majority of the plasma current within the spoke arms. While there has been much light shed on the spoke formation, structure, and effects on device performance, this brief discussion highlights that the matter is by no means settled and that this complex phenomena is worthy of continued study.

The Penning discharge is an E$\times$B device which exhibits similar behavior to that observed in Hall thrusters, albeit with a simplified geometry \cite{thomassen1966turbulent,sakawa1993excitation,raitses2015effects,rebont2011ion,kim2021magnetic}. This includes the presence of anomalous transport and rotating spokes. The device geometry, shown in Fig. \ref{fig:experiment}, consists of a central plasma source and uniform axial magnetic field. Plasma is confined axially by the dielectric end plates, and radially by the magnetic field. A radial ambipolar field emerges which drives E$\times$B processes and associated instabilities, including the spoke. The axial dynamics, along the magnetic field lines take place at a relatively short timescale when compared to the radial and azimuthal motion of the plasma. It is therefore common to assume that this axis can be decoupled from the radial-azimuthal plane. The numerical benchmark presented in this work will model a two-dimensional slice of the Penning discharge, including E$\times$B motion and formation of the rotating spoke.

As a numerical benchmark, the collisionless Penning discharge is compelling for several reasons. First, despite the complex physics taking place within the discharge, the configuration is relatively simple, and it is therefore less challenging to define a clear and concise benchmark description. Second, the discharge exhibits complex multiscale collective behavior, which should emerge from PIC code simulations. This includes microscale instabilities, such as the electron-cyclotron drift instability and magnetized two-stream instability observed in the Landmark 2a and 2b benchmarks \cite{charoy20192d,villafana20212d}, as well as large scale phenomena such as the spoke. Third, although not the original purpose of this effort, the long time-scale and non-uniform nature of the discharge strains code performance with respect to runtime and parallelisation strategy. Fourth, there exist numerous natural extensions to this benchmark, most importantly the inclusion of collisional effects. Fifth, there are several experimental Penning discharges around the world \cite{rodriguez2019boundary,pierre2004radial,kim2021magnetic}, making this, or a future collisional version of this benchmark a potential pathway towards a code validation exercise, more of which are needed within the LTP community (see for example Refs. \cite{carlsson2016validation,donko2018experimental,eremin2023modeling,eremin2024electromagnetic}). Finally, there are many fascinating and unresolved physics questions associated with partially magnetized E$\times$B discharges and the hope is that engaging numerous groups in the modeling of this device may lead to further and deeper investigation into these unsolved problems.

The paper proceeds as follows, Section \ref{sec:method} outlines the setup for the benchmark and numerical diagnostics as well as details of the participating codes. Section \ref{sec:results} compares the results of the different codes as well as highlighting lessons learned during the benchmark and providing suggestions on high-performance computing (HPC) best practices for future code development. Finally, Section \ref{sec:conclusion} offers some concluding thoughts on the effort.

\section{Methodology}
\label{sec:method}

To ensure a successful benchmarking effort, a complete, clear and concise description of the simulation configuration is crucial. In this section, we outline the simulation setup (Section \ref{sec:config}) and the diagnostics used for code comparisons (Sections \ref{sec:avg_diag} and \ref{sec:spoke_freq}). Section \ref{sec:code_details} provides details on each of the participating institutes and their associated PIC codes.

\subsection{Penning benchmark configuration}
\label{sec:config}

\begin{figure}[H]
    \centering
    \includegraphics[width=0.8\linewidth]{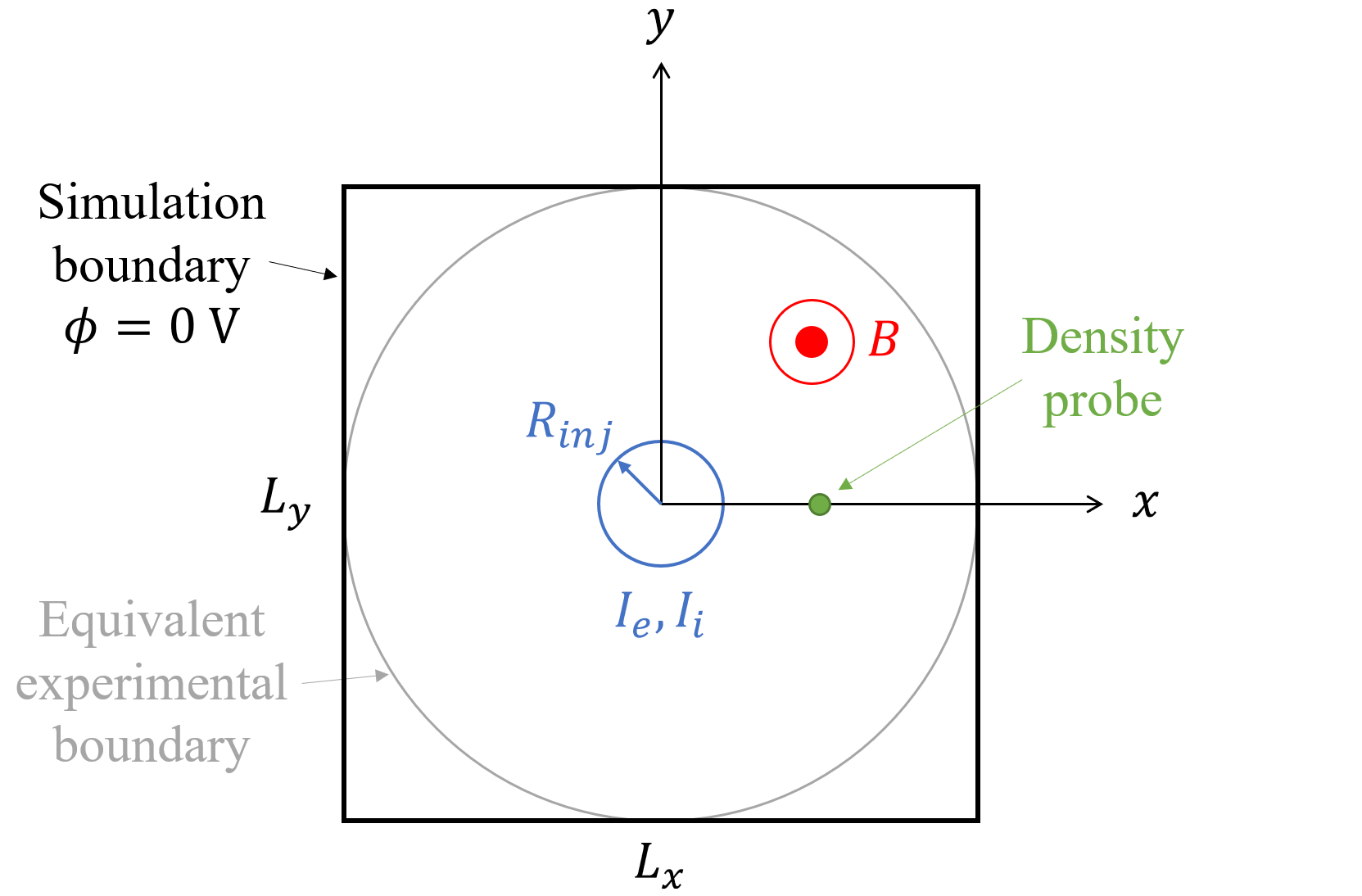}
    \caption{Schematic of the Penning discharge benchmark simulation domain, including geometry for the conducting boundaries (thick black lines), injection region (blue circle), and location of the density probe (green dot). The light gray circular boundary is not part of the simulation, but rather indicates the equivalent geometry for a Penning discharge experiment, similar to that shown in Fig. \ref{fig:experiment}.}
    \label{fig:setup}
\end{figure}

A collisionless Penning discharge configuration with helium-4 ions is considered for this benchmark. Since the spoke rotation frequency scales with the inverse square root of ion mass \cite{powis2018scaling}, typical argon and xenon gases would have led to excessive runtimes, dictating the choice of helium-4. There exists a wide range of algorithms for modeling curved boundaries in PIC simulations, each of which can lead to slightly different simulation results. To control for the effect of such implementations on the benchmarking exercise, we relax the curved geometry of the Penning discharge experiment (see Fig. \ref{fig:experiment}) to that of a grid-aligned square, an approach which has been pursued in numerous other numerical experiments \cite{chen2025three, mansour2022full, tyushev2025mode, poli2024comparison, lucken2020saturation}. The two-dimensional square domain $\{x,y\}$ is centered about the origin, with edge lengths $L_{x} = L_{y} = 5$ cm (see Fig. \ref{fig:setup}). The boundary is grounded with electric potential $\phi = 0$ V, and perfectly conducting, such that all particles reaching the boundary are absorbed (deleted from the simulation). There is a uniform, constant, externally applied magnetic field in the $z$ direction with strength $B=100$ G. With ion temperature $\approx 0.5$ eV, their cyclotron radius is comparable to the device scale, which while still making them weakly magnetized, can subtly influence the physics.

The simulation is run for a total of $N_t$ time steps each of length $\Delta t$ up to a final time $t_{\rm sim} = \SI{500}{\micro s}$. This is sufficient to resolve approximately a dozen rotations of the spoke, and $\Delta t$ is small enough to resolve the electron plasma frequency $\omega_{p,e}$ in the most dense part of the resulting quasi-steady state plasma. The grid is Cartesian and uniform with an equal number of cells in each direction ($N_{x} = N_{y}$) and with square cells ($\Delta x = \Delta y$). The cell size is sufficient to appropriately resolve the smallest electron plasma Debye length $\lambda_{D,e}$ at quasi-steady state. The values for all numerical and physical parameters are listed in Table \ref{tab:setup}. 

At $t=\SI{0}{s}$ the domain is completely empty, with no macro-particles present. A constant source of electrons and ions is created in the center of the domain, sampled randomly from a Maxwellian velocity distribution function defined by,
\begin{equation}
    f_{\alpha} \left(v_x, v_y, v_z \right) = f_{\alpha,0} \exp{\left( \frac{v_x^2 + v_y^2 + v_z^2}{2 T_{\alpha, 0}/m_{\alpha}} \right)},
    \label{eq:vdf}
\end{equation}
where $\alpha \in \{\mathrm{e},\mathrm{i}\}$ is the species index, $m_{\alpha}$ is the mass, $T_{\alpha, 0}$ is the characteristic injection temperature in Joules, and $f_{0, \alpha}$ is a normalizing constant.

The injection profile $\dot{n}_{\alpha}$ is a cylindrical Heaviside function with form,

\begin{equation}
    \dot{n}_{\alpha}(x,y)= 
\begin{cases}
    \dot{n}_{\alpha,0}& \text{for}\ x^2 + y^2 \leq R_{\mathrm{inj}}^2,\\
    0              & \text{otherwise,}
\end{cases}
\end{equation}
giving a total ``injection'' current for each species of $I_{\alpha} = \pi q_{\alpha} \dot{n}_{\alpha,0} R_{\rm inj}^2$. Where $q_{\alpha}$ is the charge and $\dot{n}_{\alpha,0}$ is the constant particle number injection rate for each species.

In these simulations, macro-particles are treated as constant weight, therefore when computing the number of macro-particles to inject per time-step, there is often a remaining fraction of a particle. This is handled by drawing a random number and testing it against the fractional remainder to determine if an additional particle should be injected per time step.

\begin{table}[H]
\centering
\label{tab:setup}
\caption{Physical and numerical parameters for the Penning discharge benchmark, including details on averaging times for diagnostics.}
\vspace{5mm}
\begin{tabular}{r|c|c|c}
\textbf{Parameter} & \textbf{Symbol} & \textbf{Value} & \textbf{Units} \\\hline \hline
helium-4 ion mass & $m_{\mathrm{i}}$ & $7291.712$ & $m_{\mathrm{e}}$ \\ \hline
Time step size & $\Delta t$ & $40$& ps\\ \hline
Number of time steps & $N_{t}$ & $12,500,000$ & - \\ \hline
Total simulation time & $t_{\rm sim}$ & $500$ & $\si{\micro s}$ \\ \hline
Steps between diagnostic print & $N_{\rm print}$ & $1,250$ & - \\ \hline
Time between diagnostic print & $\Delta t_{\rm print}$ & $0.05$ & $\si{\micro s}$ \\ \hline
Domain length in $x$/$y$-direction & $L_x$, $L_y$ & $5.0 \times 10^{-2}$ & m \\ \hline
Cells in $x$/$y$-direction & $N_{x}$, $N_{y}$ & 256 & - \\ \hline
Cell length in $x$/$y$-direction & $\Delta x$, $\Delta y$ & $1.953 \times 10^{-4}$ & m \\ \hline
Applied magnetic field strength & $B$ & $100$& G\\ \hline
Injection radius & $R_{\rm inj}$ & $5.0 \times 10^{-3}$ & m \\ \hline
Electron injection current & $I_{\mathrm{e},0}$ & $-20.0 \times 10^{-3}$ & A/m \\ \hline
Electron injection temperature & $T_{\mathrm{e},0}$ & $15.0$ & eV \\ \hline
Ion injection current & $I_{\mathrm{i},0}$ & $8.0 \times 10^{-3}$ & A/m \\ \hline
Ion injection temperature & $T_{\mathrm{i},0}$ & $0.025$ & eV \\ \hline
Macro-particle weight & $W$ & $1.0 \times 10^{5}$ & - \\ \hline
Electron macro-particles injected per time step & $N_{\Delta t, \mathrm{e}}$ & 49.932 & - \\ \hline
Ion macro-particles injected per time step & $N_{\Delta t, \mathrm{i}}$ & 19.973 & - \\ \hline
\end{tabular}
\end{table}

\subsection{Time averaged diagnostics}
\label{sec:avg_diag}

Instantaneous snapshots of parameters are printed at time-step intervals of $N_{\rm print}$ which corresponds to $\Delta t_{\rm print}$ time between outputs. The following grid-based quantities are saved at each output time:
\begin{itemize}[label=-]
    \item Ion density $n_{\mathrm{i}} \left( x,y \right)$
    \item Electron temperature $T_{\mathrm{e}} \left( x,y \right)$
    \item Plasma potential $\phi \left( x,y \right)$
\end{itemize}

Time-averaged slices of the above listed quantities are used as the primary comparison between codes. Data slices are taken along the $x$-direction with $x \in [-L_x/2,L_x/2]$ and $y = 0$ (i.e. along the $x$-axis). This single slice is then averaged over the left and right hand sides of the domain. Specifically, the data at $x \in [-L_x/2,0]$ is flipped and averaged with the data at $x \in [0,L_x/2]$, providing a single half-profile from the center of the domain to the right hand edge.

Although the system becomes quasi-neutral shortly after initialization, it takes some time for the large scale spoke structure to start-up and for the system to reach quasi-steady state. Considering this, time averaging is performed over $ \SI{200}{\micro s} \leq t <  \SI{500}{\micro s}$, with each snapshot printed at interval $\Delta t_{\rm print}$. In total, each quantity is averaged over $6,000$ instantaneous snapshots. As will be discussed in Section \ref{sec:compare_long}, several codes were run to much longer times, averaging quantities over a larger series of data.

Since the plasma potential is a grid-based quantity, measurement at each time step is straightforward. Measuring the plasma species density and temperature requires the interpolation of particle velocity moments onto the grid prior to sampling. This is performed via the standard, bi-linear, cloud-in-cell approach. Dropping the species index, the $m^{\rm th}$ velocity moment ($n U^m_{\beta}$) in direction $\beta \in \{x,y,z\}$ is computed as,

\begin{equation}
    \left( n U^m_{\beta} \right) = W \sum_{p=0}^P V_{p,\beta}^m S \left(\frac{X_p - x}{\Delta x} \right) S \left(\frac{Y_p - y}{\Delta y} \right),
\end{equation}
where $P$ is the total number of macro-particles, $X_p$ and $Y_p$ are the position coordinates of particle $p$,  $V_{p,\beta}$ is the velocity of particle $p$ in direction $\beta$, and $S(x)$ is the standard linear shape function:
\begin{equation}
    S(x) = 
    \begin{cases}
        1 + x & -1 < x < 0 \\
        1 - x & \;\;\, 0 \leq x < 1 \\
        0              & \;\;\,\text{otherwise.}
    \end{cases}
\end{equation}

The ion density is therefore computed from the $0^{th}$ velocity moment, and temperature is defined as the standard deviation of the velocity distribution function about the mean velocity. Specifically, for a given species, cell location, and instantaneous snapshot, (not indexed here), the temperature in direction $\beta$ is defined as,

\begin{equation}
    \label{eq:temp_ref}
    T_{\beta} = m \left[ \frac{\left(n U_{\beta}^2\right)}{n}  - \left( \frac{\left(n U_{\beta}\right)}{n} \right)^2 \right],
\end{equation}

For reasons that will be discussed in Section \ref{sec:compare_all}, complications arise when computing the time-average of plasma species temperature for the Penning discharge configuration. Therefore, we consider two definitions in this work.

\textbf{Temperature Type I:} The standard definition, computing the temperature using Eq.~\eqref{eq:temp_ref} at each snapshot time and then averaging temporally across all times,

\begin{equation}
    \label{eq:tempI}
    \left\langle T_{\beta} \right\rangle_t^{I} = m \left\langle \frac{\left(n U_{\beta}^2\right)}{n}  - \left( \frac{\left(n U_{\beta}\right)}{n} \right)^2 \right\rangle_t,
\end{equation}
where $\langle \cdot \rangle_t$ denotes time averaging.

\textbf{Temperature Type II:} A non-standard definition, choosing to average each velocity moment in time \textit{prior} to calculating the temperature \cite{dominguez2018particle},

\begin{equation}
    \label{eq:tempII}
    \left\langle T_{\beta} \right \rangle_t ^{II} = m \left[ \frac{\left\langle n U_{\beta}^2 \right\rangle_t}{\left \langle n \right \rangle_t}  - \left( \frac{\left\langle n U_{\beta}\right\rangle_t}{\left \langle n \right \rangle_t} \right)^2 \right].
\end{equation}

The final temperature is then computed the same for both procedures as, $\langle T \rangle_t^{\Theta} = \frac{1}{3} \sum_{\beta}\langle T_{\beta} \rangle_t^{\Theta}$ with $\Theta\in \{I,II\}$.

\subsection{Measuring spoke rotation frequency}
\label{sec:spoke_freq}

\begin{figure}[H]
    \centering
    \includegraphics[scale=0.4]{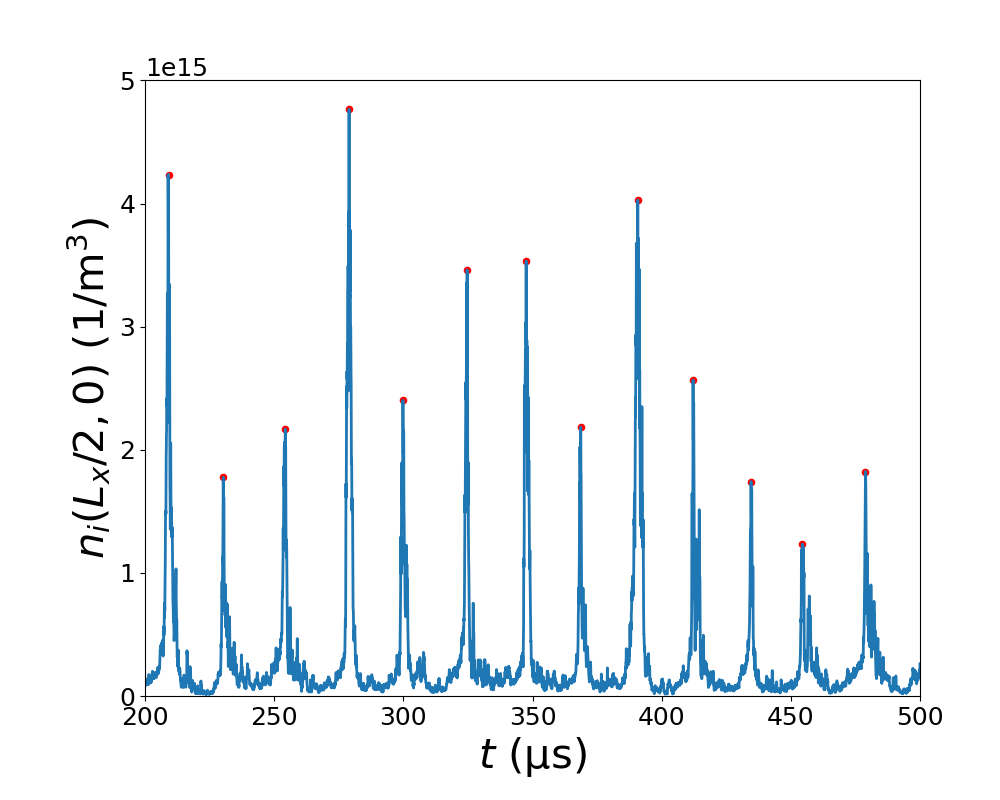}
    \caption{Plot of ion density at probe location $(x=\frac{1}{2}L_x,y=0)$ against time. Peaks (red dots) are identified via a peak-finding algorithm from the \texttt{scipy.signal} package \cite{2020SciPy-NMeth}.}
    \label{fig:spoke_freq}
\end{figure}

Many techniques could be used to measure spoke rotation frequency, however the chosen approach is both simple and reproducible. We rely on a single ion density probe (see Fig. \ref{fig:setup}) located at $(x,y)=(\frac{1}{2} L_x, 0)$, on the $x$ axis, halfway between the origin and the right hand boundary. $n_{\mathrm{i}}(\frac{1}{2}L_x,0)$ is measured at the $N_{\rm print}$ sampling rate for $ \SI{200}{\micro s} \leq t <  \SI{500}{\micro s}$, and peaks are identified via a peak-finding algorithm \cite{2020SciPy-NMeth}. To avoid detection of false peaks due to the noisy data, a minimum peak height of $n_{\mathrm{i}}=10^{15}$ $\mathrm{1/m^3}$ was set, as well as a minimum separation of 200 data points between peaks (10 \si{\micro s}). Figure \ref{fig:spoke_freq} shows a plot of density at the probe location against time and identification of the relevant density peaks. The spoke period is taken as the mean of the distance between subsequent peaks and then inverted for the frequency.

\subsection{Code details}
\label{sec:code_details}

A record forty one scientists from nineteen institutions testing seventeen codes were involved in this iteration of the \textit{Landmark} benchmarking effort. These included the Princeton Plasma Physics Laboratory - PPPL (LTP-PIC code \cite{powis2021particle}), Laplace laboratory at the University of Toulouse (EP-PIC2D \cite{Garrigues2016} and LePIC2D codes \cite{Fubiani17}), Ruhr University Bochum - RUB (RUBPIC code \cite{villafana20212d}), Wigner Research Center for Physics (MZ/X-PIC code), Dublin City University - DCU, ONERA (RHEI code \cite{portoAnisotropicElectronHeating2023}), Universidad Carlos III de Madrid - UC3M (PICASO \cite{mari24a,bell25} and PICASO-G codes), University of Saskatchewan - USask (the open-source \href{https://github.com/PrincetonUniversity/EDIPIC-2D}{EDIPIC-2D} code \cite{EDIPC-2d-web-page} - also developed with PPPL and the University of Alberta), the Laboratoire de Physique des Plasmas at \'Ecole Polytechnique - LPP (LPPic code \cite{croes2017,parodi2025}), Stanford University (XPIC2D code \cite{hara2020,hara2023}), the von Karman Institute for Fluid Dynamics - VKI (the open-source \href{https://github.com/vonkarmaninstitute/pantera-pic-dsmc}{PANTERA} code \cite{PARODI2025102244}), the European Centre for Research and Advanced Training in Scientific Computing - CERFACS (AVIP-PIC code \cite{villafana_3d_2023,bogopolsky_exploration_2025}), the Bari section of the Institute for Plasma Science and Technology of the Italian National Research Council (CNR-ISTP-Bari) with the Nuclear (NUC) Department of ENEA Frascati Research Center (PICCOLO code \cite{PICCOLO,PICCOLO2,PICCOLO3}), Sandia National Laboratories (Aleph code \cite{timko2012perform}) and the Imperial College London - ICL (IPPL-2D code \cite{faraji2023verification}).

Salient features of these codes, the computing architecture used to model the Penning benchmark, and simulation runtimes are shown in Table \ref{tab:code_details}. While most authors implemented the standard explicit momentum-conserving PIC algorithm on a structured grid, some authors considered an unstructured mesh (AVIP-PIC and Aleph codes). Codes were written in either C/C++ or Fortran, with the exception being the IPPL-2D code written with Julia. Many codes implemented external linear algebra packages for solving the Poisson equation, these packages included Hypre \cite{hypre}, PARDISO \cite{schenk2004solving}, NAG \cite{nagsolver}, PETSc \cite{petsc-efficient,petsc-user-ref,petsc-web-page} and Trilinos \cite{trilinos-website}. Some codes were parallelised on CPUs with MPI and/or OpenMP and GPUs with OpenACC or CUDA. Authors also considered different ways to decompose the problem into parallel components including particle decomposition, domain decomposition or a hybrid (mixed) approach. A deeper discussion on these approaches and their implications for performance is provided in Section \ref{sec:hpc_lessons}.


\begin{landscape}
  \begin{longtable}{cccccccc}
  \caption{Names and details of codes participating in the Penning discharge benchmark.} \label{tab:code_details}  \\
    \hline\hline
    \makecell{Institute} & Code name & Language & Poisson solver & Parallelization & Decomposition & Architecture & Wall-clock-time\\
    \hline\hline
    PPPL & LTP-PIC & C/C++ & Hypre & \makecell{MPI+OpenMP \\ w/ OpenACC} & \makecell{Hybrid \\ (particle \\ used here)} & \makecell{4x NVIDIA \\ V100 GPUs + \\ 4x IBM Power9 \\ CPU cores} & 82 hours\\
    \hline
    Laplace$^{I}$ & EP-PIC2D & Fortran & PARDISO & MPI+OpenMP & Particle & \makecell{72x Intel \\ Skylake \\ 2.30 GHz \\ CPU cores} & 71 hours\\
    \hline
    Laplace$^{V}$ & LePIC2D & Fortran & \makecell{in house \\ multigrid} & \makecell{MPI+OpenMP} & \makecell{Particle} & \makecell{28x Intel Xeon \\ CPU E5-2690 \\ v4 2.60 GHz} & 128 hours\\
    \hline
    RUB & RUBPIC & C/C++ & \makecell{NAG \\ geometric \\ multigrid} & CUDA & Hybrid & \makecell{1x NVIDIA \\ V100 GPU + \\ 1x Intel Xeon \\ Gold 5220 \\ CPU core} & 80 hours\\
    \hline
    Wigner & MZ/X-PIC & C/C++ & \makecell{in house \\ (spectral, \\ direct)} & CUDA & Particle & \makecell{1x NVIDIA \\ P100 GPU \\+ 1x Intel i7 \\ CPU core} & 10 days\\
    \hline
    DCU & DCU & C & \makecell{in house \\ multigrid} & OpenMP & Domain & \makecell{32x AMD EPYC \\ 7302 \\ CPU cores} & 10 days\\
    \hline
    ONERA & RHEI & Fortran & \makecell{in house \\ (geometric \\ multigrid)} & MPI+OpenMP & Hybrid & \makecell{48x Intel \\Sapphire \\Rapids 8468\\ CPU cores} & 16 days\\
    \hline
    UC3M & PICASO & Fortran & PARDISO & OpenMP & Particle & \makecell{40x Intel Xeon \\ Silver 4316 \\ 2.30 GHz \\ CPU cores} & \makecell{7 days \\ 13 hrs}\\
    \hline
    UC3M$^{G}$ & PICASO-G & CUDA/Python & \makecell{in house \\ (spectral, \\ direct)} & CUDA & Particle & \makecell{1x NVIDIA \\ A100 GPU } & \makecell{10 hrs}\\
    \hline
    USask & EDIPIC-2D & Fortran & \makecell{PETSc\\+Hypre} & MPI & Hybrid & \makecell{256x Intel\\E5-2683 v4 \\ 2.1 GHz \\CPU cores} & 42 days\\
    \hline
    LPP & LPPic & Fortran & Hypre & MPI & Domain & \makecell{128x AMD Rome \\ 2.6 GHz \\ CPU cores} & 55 days\\
    \hline
    Stanford & XPIC2D & C/C++ & Hypre & MPI & Particle & \makecell{32x AMD EPYC \\ 7742 2.25 GHz \\ CPU cores} & 37 days\\
    \hline
    VKI & PANTERA & Fortran & PETSc & MPI & Hybrid & \makecell{768x Intel \\ Skylake Xeon \\ Platinum 8174 \\ CPU cores} & 72 hours\\
    \hline
    CERFACS & AVIP-PIC & Fortran & \makecell{PETSc\\ (GAMG+CG)} & MPI & Domain & \makecell{192x AMD EPYC \\ 9654 (Genoa) \\ 2.4 GHz \\ CPU cores} & \makecell{7 days \\ 17 hrs}\\
    \hline
    CNR-ENEA & PICCOLO & Fortran & \makecell{PETSc + \\ hypre} & MPI+OpenMP & Hybrid & \makecell{42x AMD\\ EPYC 7402 \\ CPU cores} & 30 days\\
    \hline
    Sandia & Aleph & C/C++ & \makecell{Trilinos\\(direct solver)} & MPI & Hybrid & \makecell{640x Intel \\ Sandy Bridge \\ 2.5 GHz \\ CPU cores} & 14 days\\
    \hline
    ICL & IPPL-2D & Julia & \makecell{Julia direct \\solver} & MPI & Domain & \makecell{32x AMD EPYC \\ 2.25 GHz \\ CPU cores} & -\\
    \hline
    \hline
    \end{longtable}
\end{landscape}

\section{Results \& discussion}
\label{sec:results}

\begin{figure}[hbt!]
    \centering
    \includegraphics[width=0.9\linewidth]{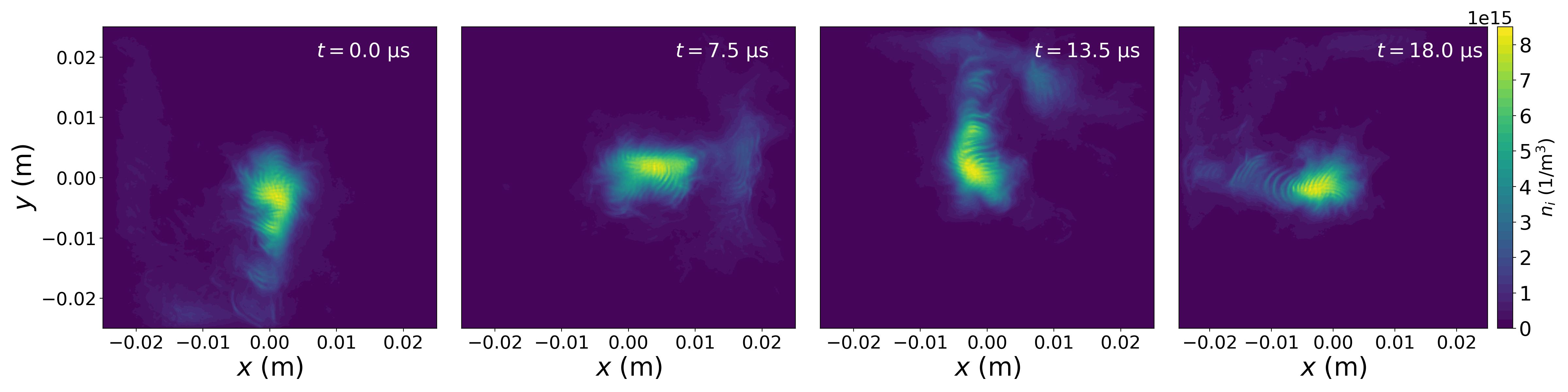}
    \caption{Contour plots of ion density at four different phases of spoke rotation after the system has reached quasi-steady state. Times are given relative to the left most figure.}
    \label{fig:spoke}
\end{figure}

Before comparing simulations across participating groups, we review a few of the key results from a single simulation. Figure \ref{fig:spoke} shows contour plots of ion density over a single period of the rotating spoke, with an animation of this figure available on \href{https://youtu.be/XWr-3vAUuvA}{YouTube} (or full URL available in Ref. \cite{spoke_movie}). Using the diagnostic described in Section \ref{sec:spoke_freq}, the mean period over 13 rotations encompassed by the \SI{300}{\micro s} averaging time is found to be  \SI{23.1}{\micro s} for a corresponding rotation frequency of \SI{43.2}{kHz}. Another important observation is that outside of the central injection region, the discharge is mostly a vacuum, except when the spoke sweeps by. In Section \ref{sec:compare_all} we will discuss how this can cause issues when comparing measurements, such as temperature, between codes.

As discussed in Section \ref{sec:intro} spoke formation may be the result of various mechanisms; however, due to the uniform magnetic field and collisionless nature of these simulations, it is hypothesized that the spoke shown here is a saturation of the collisionless Simon-Hoh instability (a type of GDI). Under this assumption, Eq. 4 in Ref. \cite{powis2018scaling} provides a formula for the spoke rotation frequency as a function of the discharge radius, ion mass, radial electric field, and gradient length scale. For this simulation we have a radial electric field of approximately $100$ V/m and gradient length scale of $7.1$ mm giving a predicted spoke frequency of $53$ kHz, which is reasonable compared to the measured frequency.

Figure \ref{fig:spoke}, and associated animation, also demonstrate the complex, chaotic and multiscale nature of the discharge. On the largest scales we clearly have the rotation of the relatively slowly evolving spoke; however, this rotation does not appear to be rigid, and the animation in particular shows the persistent shedding of density at various phases of spoke rotation. Although harder to observe, there are also finer scale structures which exist within the core of the discharge and along the spoke arm, which may be a form of ion-acoustic or electron-cyclotron-drift instabilities \cite{forslund1970electron,janhunen2018evolution,charoy2020comparison,tyushev2025mode}. Since this simulation is collisionless, radial transport of magnetized electrons can only be induced by these waves and instabilities and is highly correlated with the spoke rotation. A further discussion or analysis of this transport is beyond the scope of this benchmarking paper, however we refer the reader to many of the excellent references in Section \ref{sec:intro} which explore this in depth.

\subsection{Long time scale simulation comparison}
\label{sec:compare_long}

The first simulation comparisons are made between codes which were run for extended times, with the goal of collecting uncertainty bounds for time averaged diagnostics. The list of codes, run times, and associated measured spoke frequencies are shown in Table \ref{tab:long_sims}. From $0.2-4.0$ ms the probe in the PPPL code measures 163 passages of the spoke for a, previously mentioned, mean frequency of $43.2$ kHz. The uncertainty bounds are defined by computing averaged frequencies over each group of 13 sequential spoke rotations and then considering the maximum and minimum frequencies from the set. This gives a min/max mean frequency of $41.1$ kHz and $46.1$ kHz, respectively. Both Laplace codes (EP-PIC2D and LePIC2D) agree with this value within the uncertainty bounds.

\begin{table}[H]
\centering
  \caption{Details for long time simulations including spoke frequency measurements.}
  \label{tab:long_sims}
  \begin{tabular}{cccll}
    \hline
    Institute & Code name & Total time & \makecell[l]{Spoke\\ frequency} & \makecell[l]{Frequency\\discrepancy}\\
    \hline
    PPPL & LTP-PIC & $4.0$ ms & 43.2 $\left( \begin{array}{c} +2.9 \\ -2.1 \end{array} \right)$ kHz & -\\
    Laplace$^I$ & EP-PIC2D & $5.0$ ms & $43.3$ kHz & $0.1$ kHz\\
    Laplace$^V$ & LePIC2D & $1.6$ ms & $42.6$ kHz & $0.8$ kHz\\
    \hline
\end{tabular}
\end{table}

\begin{figure}[H]
    \centering
    \includegraphics[width=0.5\linewidth]{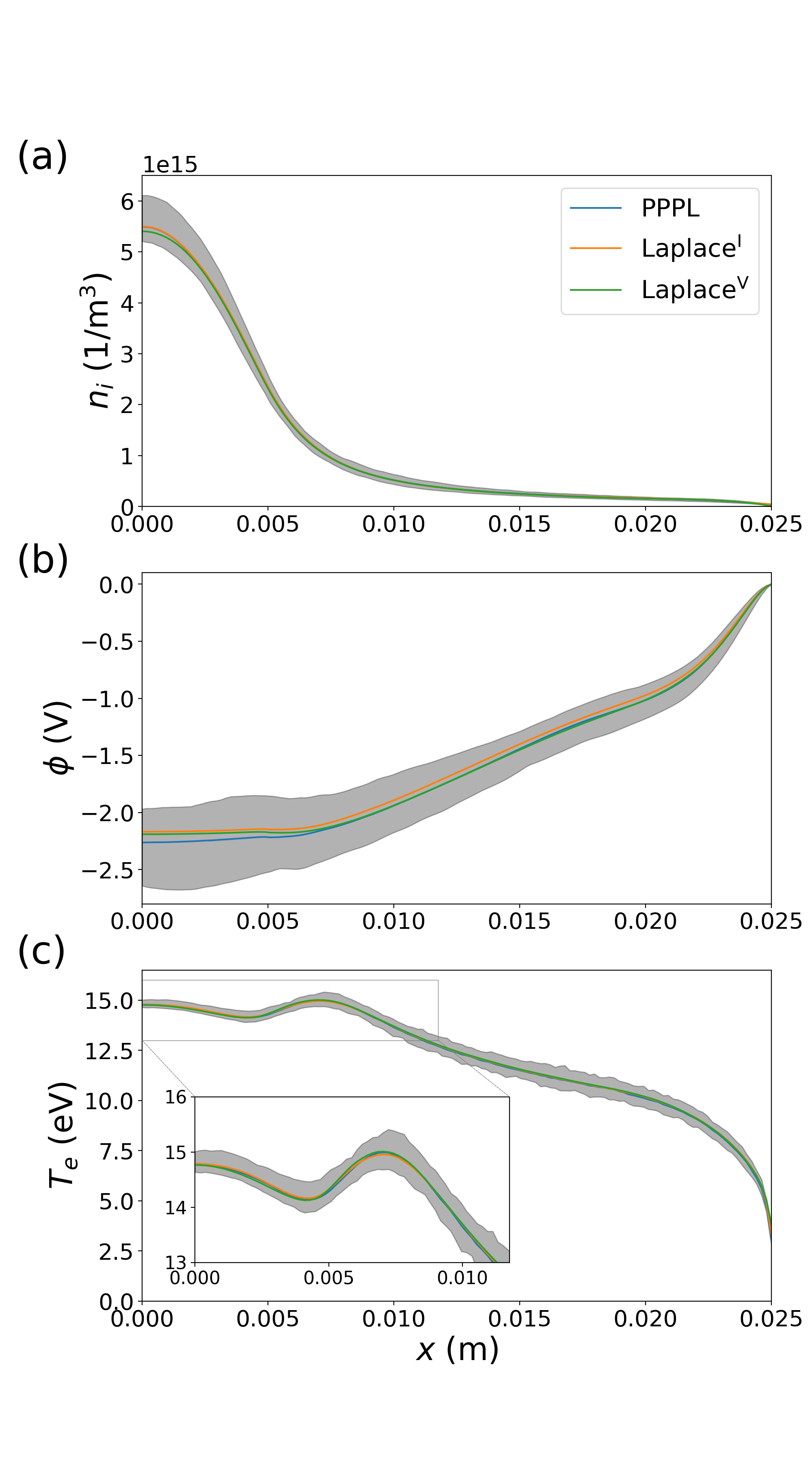}
    \caption{Comparison of time-averaged slices for long-time simulations: (a) ion-density, (b) plasma potential, and (c) electron temperature (Type I).}
    \label{fig:compare_long}
\end{figure}

Figure \ref{fig:compare_long} compares time averaged half-profile slices for (a) ion density, (b) electric potential, and (c) electron temperature (Type I). The uncertainty bounds (shown as gray in the figure) are obtained by considering statistics of a rolling average, over \SI{300}{\micro s}, of the PPPL data only. Each average begins at the start of the subsequent snapshot, providing a total of 7,000 averages. The lower and upper bounds are computed as the minimum and maximum values within this set of averages at each grid location. This definition proved critical to assessing agreement within the larger comparison between all codes, as will be discussed in Section \ref{sec:compare_all}. Using this uncertainty bounds, good agreement is observed between all three codes considered here.

\subsection{Simulation comparison for all codes}
\label{sec:compare_all}

\begin{figure}[H]
    \centering
    \includegraphics[width=0.6\linewidth]{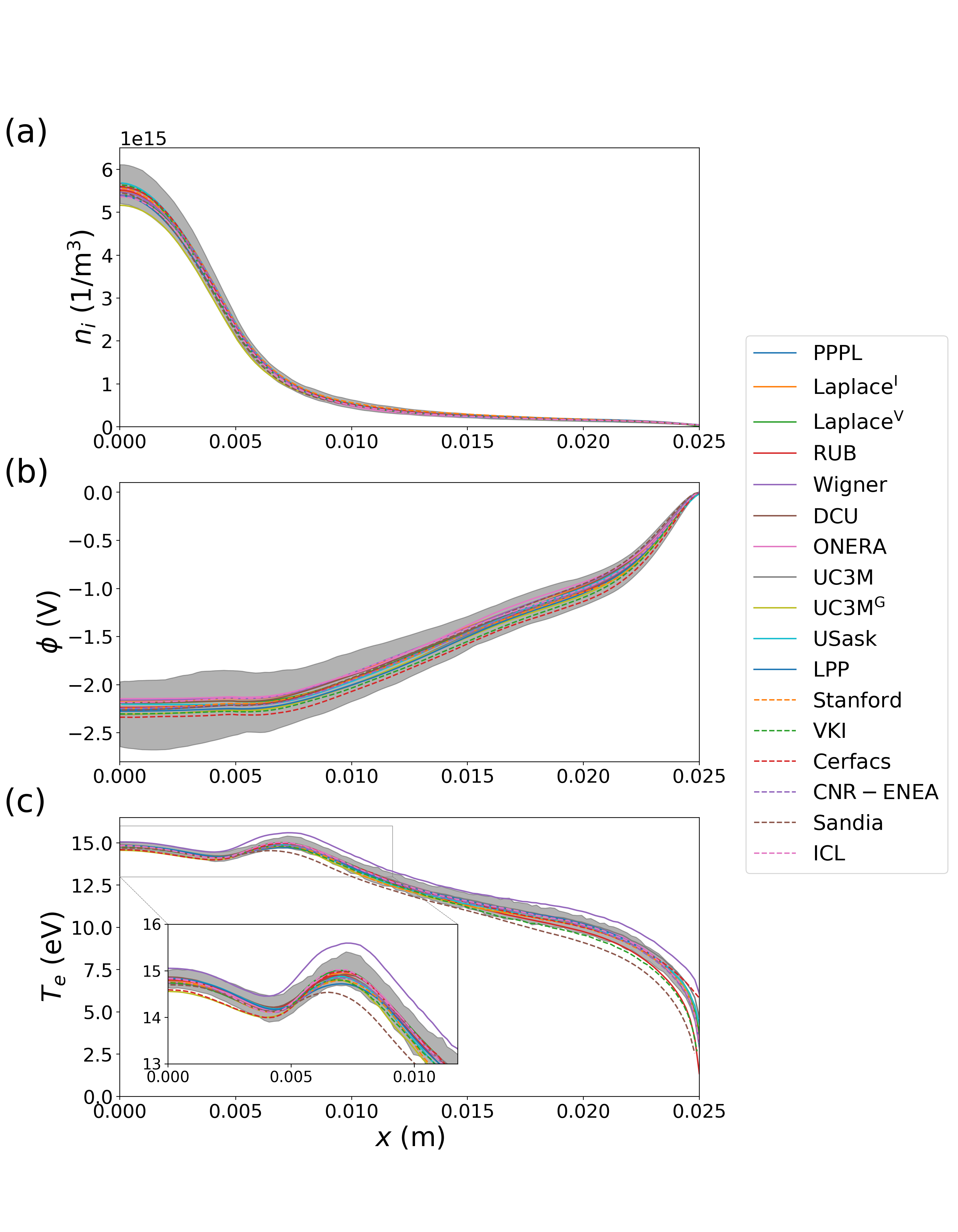}
    \caption{Comparison of time-averaged slices between all codes: (a) ion density, (b) plasma potential, and (c) electron temperature (Type I). See supplementary material for the mean and uncertainty bounds data for each plot.}
    \label{fig:compare_all}
\end{figure}

Proceeding with a comparison between all codes, Fig. \ref{fig:compare_all} shows plots of time averaged half-profile slices for (a) ion density, (b) electric potential, and (c) electron temperature (Type I). Note that the results from Section \ref{sec:compare_long} for long time scale simulations are reproduced here for the relevant codes. Using the same uncertainty bounds defined in Section \ref{sec:compare_long}, all codes agree for plots of ion density and plasma potential. Turning to the electron temperature, we observe mixed results, with most codes falling within the tight uncertainty bounds, but with several lying outside of these bounds either at the center or edges of the discharge.

Understanding the cause of this discrepancy in electron temperature across simulations resulted in significant further analysis and a large number of additional simulations. Fundamentally, there are two issues with the benchmark which resulted in this outcome. The first is that the system is highly multiscale, especially with the presence of the low-frequency rotating spoke. Therefore, the proposed \SI{300}{\micro s} averaging time is likely too short to achieve convergence between simulations. With consideration to the simulation run times listed in Table \ref{tab:code_details}, it was decided that asking all groups to simulate longer times to improve averaging would be prohibitive. Therefore, this issue was remediated by having a few select groups perform very long time simulations to; (1) check that convergence could be achieved at these time scales and (2) define the uncertainty bounds which could be used to encapsulate accuracy from all codes, the results of which were discussed in Section \ref{sec:compare_long}. This demonstrated excellent agreement between the three independently developed codes (from PPPL, and two from Laplace) across all tested diagnostics, suggesting that if all codes were run to this length better agreement would be observed.

The second issue arises due to the persistence of vacuum regions throughout the simulation, as shown in Fig. \ref{fig:spoke}. When capturing snapshots of temperature data there can be ambiguity in how to handle cells with few or zero macro-particles. When there are zero macro-particles in a given cell, most groups set the temperature to zero, whereas in reality the temperature is undefined. When averaging over time, these points should be excluded from the procedure, however most codes include the data point as a zero. This would ordinarily not be a problem since most PIC simulations fill the area with particles, however the vacuum regions in this configuration exacerbate the issue. Things become more subtle when considering cells with few, or a fraction of a particle (which is possible for cloud-in-cell shape functions). In particular, if there are only a few particles, the statistics for sampling temperature are poor, leading to significant noise. This noise can lead to anomalously high temperature calculations within a given cell which can easily skew the time averaged results. Different codes also computed temperature in slightly different ways, with some using sample variance and others relying on population variance, which can be related to each other via Bessel's correction \cite{reichmann1962use}. For a well-sampled distribution, these values would agree closely, however in the vacuum region, which are poorly sampled by particles, the discrepancy can be large.

Further issues included whether codes were defined with boundaries at cell-centers or cell-nodes, the type and convergence tolerance of the numerical methods used to solve the Poisson equation, and the effect of the Pseudo-Random-Number-Generator algorithm. However, it is worth noting that sensitivity tests with the PPPL and the two Laplace codes revealed that neither of these factors played a strong role in influencing the temperature profiles.

\begin{figure}[H]
    \centering
    \includegraphics[width=0.6\linewidth]{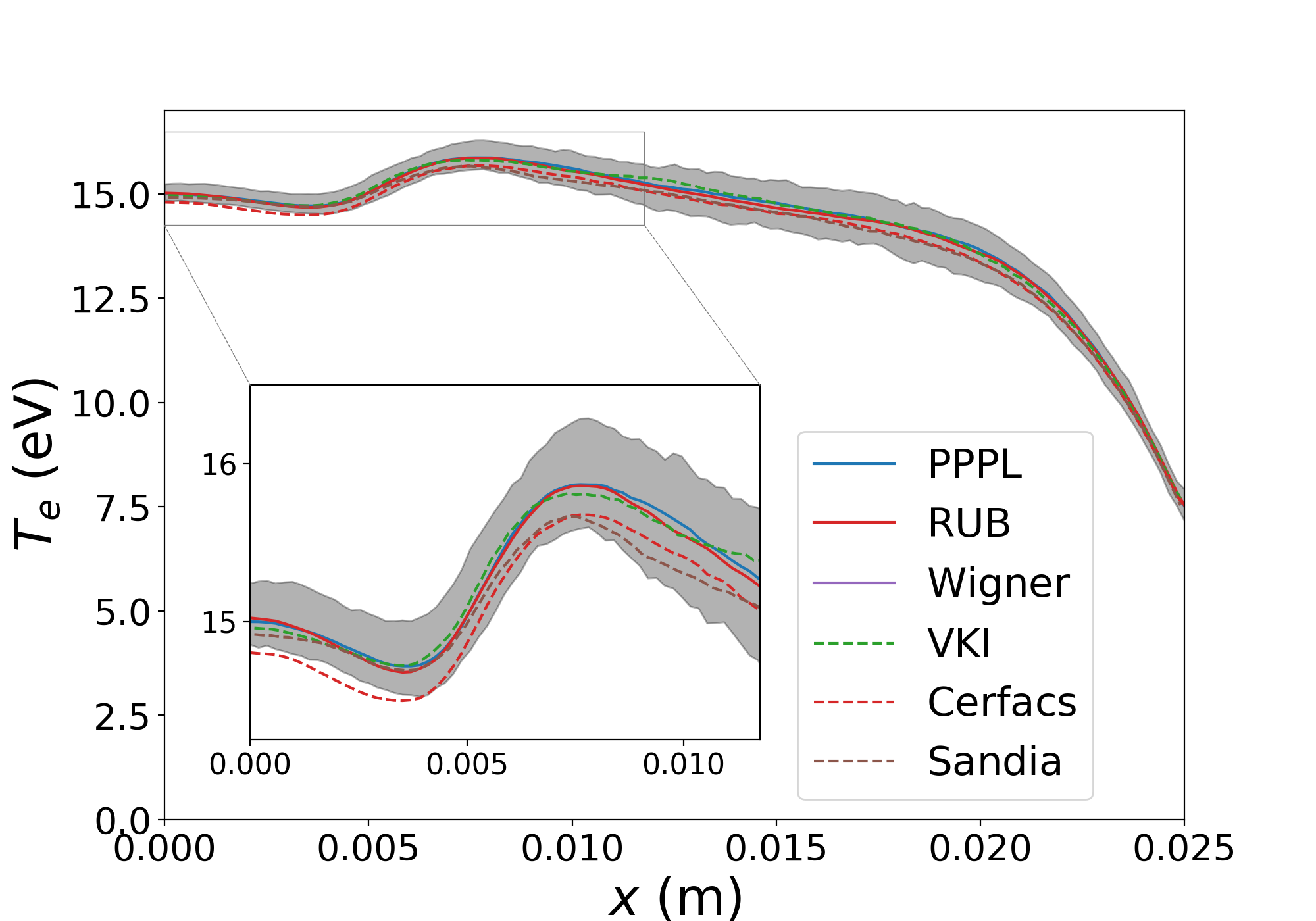}
    \caption{Comparison of time-averaged slices of electron temperature (Type II) for select codes. See supplementary material for the mean and uncertainty bounds data of the plot.}
    \label{fig:compare_alt_temp}
\end{figure}

To remedy this discrepancy between codes, we asked several of the ``outlier'' code authors to rerun their simulations with the alternative temperature diagnostic described by Eq.~\eqref{eq:tempII} (\textbf{Temperature Type II}). Averaging velocity moments prior to calculating temperature has several advantages. This includes that velocity moments are truly zero when there are no particles in the cell, and time-averaged moments gather significant statistics to end up being non-zero everywhere (except at the boundaries). Furthermore, it reduces the chance of small values being present in the denominator of the temperature calculation, further improving statistical accuracy. While this is not strictly the standard definition of time-averaged temperature, it is a valid way to compare numerical methods if done consistently. Figure \ref{fig:compare_alt_temp} shows plots of \textbf{Temperature Type II} computed via Eq.~\eqref{eq:tempII}, demonstrating agreement between nearly all previously outlier codes within the uncertainty bounds.

While investigating the causes of discrepancy between groups, numerous simulations were run with both the Sandia and CERFACS codes in an attempt to understand how the unstructured nature of their meshes could influence the results of the benchmark. Although no clear influence was identified, suspected impacts include that (1) the location of the triangular cell center may not align with those from a Cartesian grid, and (2) the smaller area of the triangular cell may lead to an increase in measured fluctuations in regions of low statistical resolution. The Aleph code resolved outstanding discrepancies by printing complete particle phase space data at each time step and mapping this data onto a Cartesian grid in post-processing. Despite best attempts, the Cerfacs code still fell around 1.3\% outside of the Temperature Type II measurement uncertainty bounds near the center of the simulation. We suspect this is due to the unstructured nature of the code since it has been well benchmarked in previous Landmark efforts \cite{charoy20192d,villafana20212d}.

The newly developed UC3M$\mathrm{^G}$ code also showed some minor discrepancies close to the uncertainty bounds for the density and temperature (Type I) measurements. After extensive investigation, this discrepancy was conclusively attributed to the group's use of a spectral field solver, which showed root-mean-square fluctuations 3-15\% in the electric field larger than those from a finite-difference field solver. 

\begin{figure}[H]
    \centering
    \includegraphics[width=0.6\linewidth]{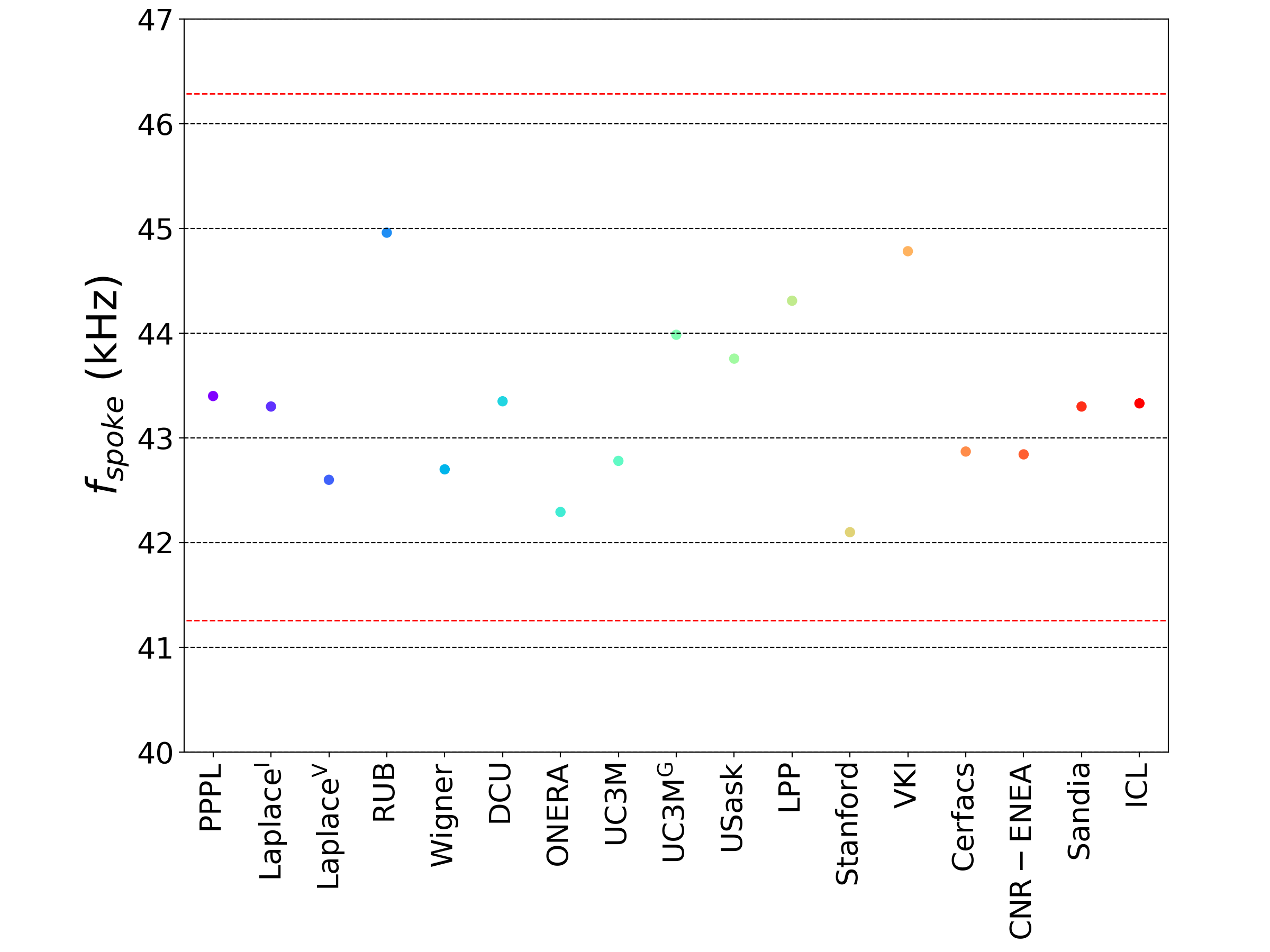}
    \caption{Comparison of the spoke rotation frequency, showing agreement between all codes. The reference frequency of $43.2$ kHz is produced from long time simulations with the PPPL code with uncertainty bounds of $-2.1$ kHz and $+2.9$ kHz delineated by the red dashed lines (see Section \ref{sec:compare_long} for further details).}
    \label{fig:compare_all_spoke}
\end{figure}

Finally, Fig. \ref{fig:compare_all_spoke} compares the spoke rotation frequency measured by all groups, showing agreement within the measured frequency bounds of $[41.1,46.1]$ kHz from long time simulations using the PPPL code. Despite the challenges associated with this benchmark, we believe that the results presented in this section constitute sufficient evidence to conclude that there is agreement between almost all codes across a variety of diagnostics.

\subsection{Lessons learned for benchmark selection}

Despite the best efforts of the authors, there were several significant challenges associated with completing this benchmarking comparison, some of which are discussed in Sec. \ref{sec:compare_all}. In this section, we aim to highlight some of the lessons learned from this process with the hope that future benchmarking efforts can avoid the same pitfalls.

\textbf{Benchmark lesson 1: Prepare a well defined benchmark description.} Based on the author's experience participating in previous PIC benchmarks, a best attempt was made to prepare a clear, concise, and sufficient description of the Penning discharge setup as well as the diagnostics used to compare codes. Despite this effort, several important details were missing, which in hindsight would have proved useful. These included an appropriate definition of how to handle temperature in vacuum regions and the specific equations for calculating radial injection locations on the Cartesian grid. The choice of the pseudo-random number generator was also a question raised at code author meetings, although it was ultimately demonstrated that this had little influence on the results. Nonetheless, we emphasize that preparing a suitable benchmark description is paramount to success in any such effort.

\begin{figure}[H]
    \centering
    \includegraphics[width=0.6\linewidth]{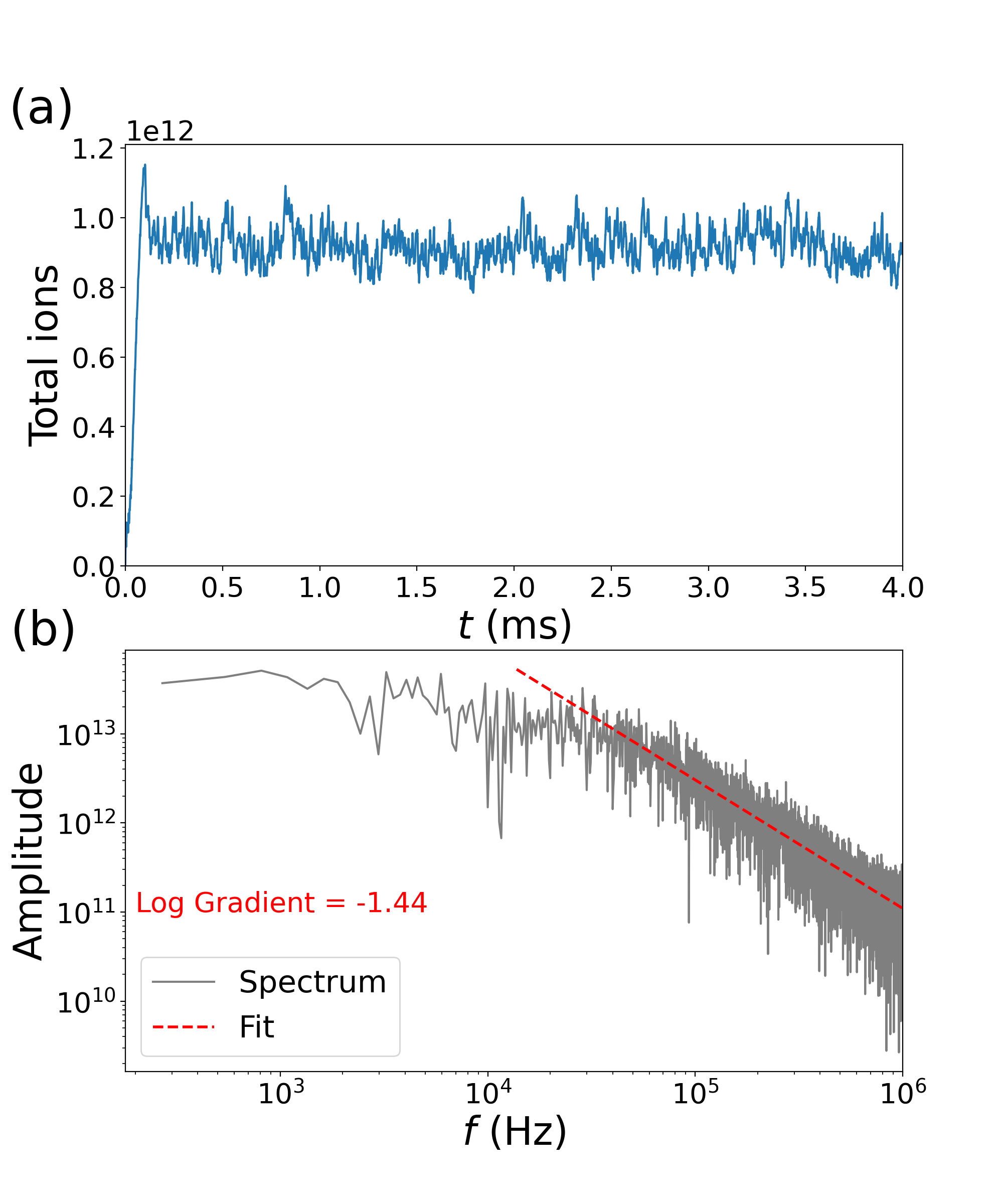}
    \caption{(a) Plot of total number of ions within the simulation vs time, and (b) frequency spectrum of the data from plot (a), demonstrating the chaotic nature of the Penning discharge benchmark configuration.}
    \label{fig:spectrum}
\end{figure}

\textbf{Benchmark lesson 2: Pick a low cost benchmark.} A major challenge associated with obtaining convergence between all codes was the large simulation cost, both in computing resources and time, of modeling the Penning discharge (see Table \ref{tab:code_details}). This made it challenging for many groups to rerun their simulations or run for longer times, which proved essential for demonstrating agreement. Of course, a benchmark should run for sufficient time such that it tests the capabilities of each code to reproduce important behavior relevant to the domain of physics, but ideally it should not be longer than that. This is particularly challenging in the LTP community, where the phenomena under investigation are often highly multi-scale, especially when collective effects and collisions are included. Figure \ref{fig:spectrum} shows how the spectrum of the total number of ions in the simulation follows a power law, indicative of chaotic behavior \cite{pope2001turbulent} and the highly multiscale nature of the discharge, making it less ideal as a benchmark configuration. In conclusion, a benchmark should be selected that is long enough to capture the physics being tested, but ideally not longer.

\textbf{Benchmark lesson 3: Avoid systems which have regions of poor statistic resolution.} This lesson pertains particularly to particle-based methods, such as the PIC algorithm, which is an approximate solution to the Vlasov equation. The discussion in Section \ref{sec:compare_all} highlights that poor statistical sampling within the vacuum regions of the Penning discharge led to significant discrepancies in temperature measurements between the codes. In general, poor statistical sampling can lead to high, and unpredictable levels of noise which can easily skew results. It is therefore recommended to avoid simulations which consist of these poorly resolved regions. As an example, the \textit{Landmark} 2a \cite{charoy20192d} and 2b \cite{villafana20212d} benchmark configurations provide a more appropriate setup which avoids these pitfalls.

\textbf{Benchmark lesson 4: More data is better.} Even with a well prepared benchmark description, it is impossible for the researchers to predict all possible ways in which disagreement may arise between codes. To prepare for this eventuality, it is wise to request that participants collect more than the minimum data required to compare results. Case in point, if participants in this benchmarking effort had been requested to record velocity moments, rather than temperature, it would have allowed for a rapid comparison between the Type I and Type II temperature diagnostics, eliminating the need for further simulations. Additionally, collecting more complete data can make it easier to determine uncertainty bounds for each simulation, which can further simplify comparison.

Despite the significant time that went into resolving issues with this benchmark, the authors want to emphasize that such efforts will continue to be critical for ensuring that complex codes can accurately reproduce results which are relevant to the broader community. More complex benchmark configurations, such as those with collisions, are also required, and the ultimate goal of the community should be to perform rigorous validation studies against experimental data.  The lessons learned here will be essential to achieving that goal.

\subsection{High performance computing lessons learned}
\label{sec:hpc_lessons}

Although the original intent of this benchmarking effort was not to assess or even stress test the computational performance of the participating codes, the results shown in Table \ref{tab:code_details} (comparing code implementations and performance) reveal some interesting trends that pertain to code implementation and high-performance computing (HPC) strategies.

\textbf{HPC Lesson 1: Parallel decomposition strategy matters.} One trend which became apparent throughout the benchmark is the total runtime for a given code compared to the parallel decomposition strategy. For PIC simulations, there are generally two approaches. The simplest, particle decomposition, is where the list of particles in memory is split (usually evenly) between MPI tasks, each of which owns a copy of the entire simulation grid. Particle decomposition is simple to implement and easy to load balance, however the downside of each task owning an entire copy of the grid is the requirement for at least two \texttt{MPI\_Allreduce} calls at each time step to share grid information. Therefore, most codes implement a domain decomposition approach, whereby each task owns a subset of the global mesh and all particles which reside on that mesh. This avoids the need for an expensive global MPI call, and improves scaling performance, however introduces code complexity due to particle communication across domain boundaries.

The Penning discharge configuration of this benchmark also highlights the second challenge with domain decomposition, that of load balancing. As shown in Fig. \ref{fig:spoke}, the spoke, and associated simulation particles, comprise only a narrow angular region of the domain outside of the central injection region. Therefore, at any given time, a uniformly domain decomposed grid will have extremely poor load balancing, effectively restricting code performance to that of a few cores handling a majority of the particles. By comparing the ``Decomposition'' and ``Runtime'' columns of Table \ref{tab:code_details}, it is clear that codes which implement particle decomposition performed, on average, better than those with domain decomposition. We note that the situation may improve for collisional simulations, where scattering can better distribute plasma density throughout the domain.

Despite this trend, the value of domain decomposition cannot be understated. For large simulations, particle decomposition becomes increasingly expensive and eventually intractable due to communication bottlenecks and memory limitations. However, the lesson learned here is that a hybrid approach to parallel decomposition can improve versatility of code performance across a wide range of applications. Such an approach has two levels of task parallelism, with the outer (higher) level being domain decomposition and then the inner (lower) level being particle decomposition within each domain. This flexibility allows a code to better handle load balancing when a system has large plasma gradients, yet still achieve reasonable parallel scaling. Alternatively, one could consider dynamic load balancing \cite{germaschewski2016plasma,miller2021dynamic,nakashima2009ohhelp} where the domain owned by each process is adjusted on the fly based on the spatial density of the simulation particles.

\textbf{HPC Lesson 2: GPUs are effective at accelerating PIC codes.} A second trend which emerges from Table \ref{tab:code_details} relates to the hardware targeted by each code. In general, codes which are able to take advantage of GPU acceleration perform faster than those using CPUs. In the last decade, general purpose graphical processing units have provided an affordable, massively parallel alternative to multi-core CPUs. At the time of writing, these chips have a core count on the order of 10,000, and thread counts over 1 million for each GPU device, making them ideal for acting on large chunks of contiguous memory, such as particle phase-space data. Furthermore, the implementation of efficient atomic operations in languages such as CUDA (which can be targeted by abstraction layers such as OpenMP and OpenACC) leads to very fast interpolation routines. GPUs also benefit from large caches of high speed memory, which can store the entire local grid chunk, further accelerating interpolation and grid based routines. Details on writing low-temperature plasma PIC codes for GPUs can be found in Refs \cite{juhasz2021efficient,powis2021particle,jambunathan2018chaos}.

\textbf{HPC Lesson 3: Adhering to HPC best practices is essential.} Even when optimizing over parallel strategies and chip architectures, it is clear from Table \ref{tab:code_details} that an enormous amount of time is required for each Penning discharge simulation. This results from a familiar problem to all computational plasma physicists, the vast range of time scales exhibited by plasma phenomena, which are particularly present in simulations with low-frequency spoke oscillations. Although in the previous section, we recommended avoiding such long time scales in future benchmarks, this obviously does not eliminate the need to model such waves and instabilities for future scientific and engineering purposes. If groups are to continue relying on the standard explicit PIC method, then high-performance computing best practices are essential to reduce the computing time per simulation step, and therefore total simulation time. Details on how to achieve these have been documented extensively \cite{kaganovich2020physics,juhasz2021efficient,taccogna2023plasma,powis2021particle}.

This lesson also extends to selecting external packages which will be integrated into a code. For LTP simulations, this is usually in the form of the linear algebra package chosen (or written in-house) to solve the Poisson equation. The choice of solver can play an important role, with direct solvers being best for small domains and Krylov and/or multigrid methods being best for larger domains \cite{saad2003iterative}.

Broadly, this lessons come down to understanding how to efficiently map the fundamental PIC algorithms onto relevant (yet constantly changing) computational architecture for maximum performance. With a trend towards increasing heterogeneity in computer systems, this requirement will only continue to grow in importance over the coming years. 

\textbf{HPC Lesson 4: Consider alternative algorithms.} Code developers could also consider alternative algorithms which can step over the usually strict numerical resolution requirements of PIC, these include energy-conserving PIC, which allow for larger cell sizes \cite{lewis1970energy,barnes2021finite,powis2024accuracy}, as well as implicit \cite{lapenta2017exactly,chen2011energy,eremin2022energy} or semi-implicit \cite{langdon1983direct,gibbons1995darwin,sun2023direct} time-stepping techniques, allowing for larger time steps. Advanced PIC algorithms, which may offer further performance improvements, include sparse-grid PIC \cite{deluzet2022sparse,garrigues2021application1,garrigues2021application,garrigues2024acceleration1,garrigues2024acceleration} or reduced-order PIC \cite{faraji2022enhancing,reza2023concept,reza2024latest}, both of which have proven accurate for modeling E$\times$B discharges. Of course, for accurate simulation results, care must always be taken to resolve the appropriate time and length scales of the relevant physical phenomena which give rise to the spoke and associated anomalous transport.

We also wish to highlight some of the expected performance challenges associated with the unstructured Sandia and CERFACS codes. In general, the handling of particle data in memory, and transfer of data between adjacent cells can incur an increased overhead. Furthermore, when particle phase data is stored contiguously there is an increased cost to identifying the parent cell, which is otherwise a trivial process on a Cartesian grid. Finally, unstructured meshes are unable to take advantage of the highest performing linear algebra preconditioners such as geometric multigrid. Although an unstructured grid can lead to increased simulation cost, the benefit is significantly increased flexibility, especially when modeling complex engineering systems.

\section{Conclusion}
\label{sec:conclusion}

The computational study of low-temperature plasmas is essential to future progress in plasma science and engineering. A critical part of this effort is the verification, or benchmarking, of codes used to simulate this phenomenon. The two-dimensional Penning discharge configuration detailed in this article has many attributes which make it an ideal stress test of the required physics modeling and computational performance of low-temperature plasma kinetic simulation software. Unlike the \textit{Landmark 2} series of benchmarks, this new configuration spans an order of magnitude larger range of time scales, and captures the emergence of the large scale rotating spoke structure. Seventeen codes, a record for this community, were put to the test on this benchmark, with agreement found for nearly all across four important measurements.

Several challenges were encountered throughout this process, resulting from a lack of detail in the benchmark description, high computational cost of the configuration, and poor statistical resolution in regions of the discharge. The effort also highlighted important computational considerations for particle-in-cell code design, including flexibility of parallelization strategy, adherence to high-performance computing best practices, and taking advantage of modern computing hardware. The authors hope that describing these lessons will help this, and other simulation communities, design improved benchmarks for their respective software.

Despite a growing number of benchmarks within this community, new configurations are required to test more complex low-temperature plasma physics, including collisions, chemical reactions, and surface processes. Furthermore, validation against experimental data is severely lacking, and essential to the future use of these codes for modeling complex engineering systems. We hope that this effort serves as a small step along the road towards developing robust, efficient, and useful plasma simulation software.

\section*{Acknowledgments}

This research was supported by the U.S. Department of Energy through the PPPL CRADA agreement with Applied Materials, the Princeton Collaborative Research Facility (PCRF) under contract No. DE-AC02-09CH11466, and the Laboratory Directed Research and Development (LDRD) program. The United States Government retains a non-exclusive, paid-up, irrevocable, world-wide license to publish or reproduce the published form of this manuscript, or allow others to do so, for United States Government purpose(s).

The work of Federico Petronio was partially supported by Agence de l’Innovation de Defense – AID - via Centre Interdisciplinaire d’Etudes pour la D\'efense et la S\'ecurit\'e – CIEDS - (project 2023 – validHETion). This project was provided with computer and storage resources by GENCI at TGCC thanks to the grant 2024-A0160510439 on the supercomputer Joliot Curie's Irene-ROME partition.

For what concerns the CNR-ENEA study,
simulations with the PICCOLO code were carried out with Eni HPC4 Supercomputing Cluster, under the Joint Research Agreement ENI-CNR on 'Fusione a Confinamento Magnetico'.


The Wigner Research Center for Physics would like to thank the Hungarian Office for Research, Development and Innovations for their support through grant number 152843.

This work was supported by DOE grant No. DE-SC0022201. This work used the capabilities of the SNL Plasma Research Facility, supported by DOE SC FES. This article has been authored by an employee of National Technology \& Engineering Solutions of Sandia, LLC under Contract No. DE-NA0003525 with the U.S. Department of Energy (DOE). The employee owns all right, title and interest in and to the article and is solely responsible for its contents. The United States Government retains and the publisher, by accepting the article for publication, acknowledges that the United States Government retains a non-exclusive, paid-up, irrevocable, world-wide license to publish or reproduce the published form of this article or allow others to do so, for United States Government purposes. The DOE will provide public access to these results of federally sponsored research in accordance with the DOE Public Access Plan https://www.energy.gov/downloads/doe-public-access-plan.

The work of Pietro Parodi is funded by an FWO Strategic Basic PhD fellowship (reference 1S24022N). The authors of the VKI group gratefully acknowledge the Gauss Centre for Supercomputing for providing computing time on SuperMUC-NG at Leibniz Supercomputing Centre through the project ``Heat flux regulation by  collisionless processes in heliospheric plasmas – ARIEL''.

This work conducted by the Ruhr University Bochum group was funded by the German Research Foundation within the framework of the project ``Modellierung und Simulation von Hochleistungsmagnetronentladungen'' (No. 550860775).

The contribution of E. Ahedo, M.P. Encinar, P. Fajardo, A. Marín-Cebrián has been supported by the R\&D project PID2022-140035OB-I00 (HEEP) funded by MCIN/AEI/10.13039/501100011033 and by ``ERDF A way of making Europe''.
The contribution of E. Bello-Benítez, M. Merino has received funding from the European Research Council (ERC) under the European Union’s Horizon 2020 research and innovation programme (Starting Grant ZARATHUSTRA, grant agreement No 950466).

This work conducted at Laplace was granted access to the HPC resources of CALMIP supercomputing center under Allocation No. 2013-P1125.

G. Bogopolsky acknowledges financial support from Safran Spacecraft Propulsion and the Association Nationale de la Recherche et de la Technologie (ANRT) as part of a CIFRE convention.

\bibliographystyle{ieeetr}
\bibliography{sample}

\end{document}